\documentstyle[epsf,eqsecnum,floats,preprint,aps]{revtex}

\tighten

\input epsf
\begin{document}
\newcommand{\beq}{\begin{equation}}
\newcommand{\eeq}{\end{equation}}

\newcommand{\be}{\begin{equation}}
\newcommand{\ee}{\end{equation}}
\newcommand{\bea}{\begin{eqnarray}}
\newcommand{\eea}{\end{eqnarray}}
\newcommand{\PSbox}[3]{\mbox{\rule{0in}{#3}\includegraphics{#1}\hspace{#2}}}
\overfullrule=0pt
\def\Int{\int_{r_H}^\infty}
\def\d{\partial}
\def\e{\epsilon}
\def\M{{\cal M}}
\def\high{\vphantom{\Biggl(}\displaystyle}
\catcode`@=11
\def\@versim#1#2{\lower.7\p@\vbox{\baselineskip\z@skip\lineskip-.5\p@
    \ialign{$\m@th#1\hfil##\hfil$\crcr#2\crcr\sim\crcr}}}
\def\simge{\mathrel{\mathpalette\@versim>}} %
\def\simle{\mathrel{\mathpalette\@versim<}} %
\catcode`@=12 

\def\pr#1#2#3#4{Phys. Rev. D {\bf #1}, #2 (19#3#4)}
\def\prl#1#2#3#4{Phys. Rev. Lett. {\bf #1}, #2 (19#3#4)}
\def\prold#1#2#3#4{Phys. Rev. {\bf #1}, #2 (19#3#4)}
\def\np#1#2#3#4{Nucl. Phys. {\bf B#1}, #2 (19#3#4)}
\def\pl#1#2#3#4{Phys. Lett. {\bf #1B}, #2 (19#3#4)}

\rightline{CU-TP-964}
\rightline{hep-th/0001140}
\vskip 1cm

\begin{center}
\ \\
\large{{\bf Gravitational properties of monopole spacetimes\\
near the black hole threshold}}
\ \\
\ \\
\ \\
\normalsize{Arthur Lue\footnote{\tt lue@phys.columbia.edu} and 
Erick J. Weinberg\footnote{\tt ejw@phys.columbia.edu}}
\ \\
\ \\
\small{\em Department of Physics \\
Columbia University \\
New York, NY 10027}

\end{center}

\begin{abstract}

Although nonsingular spacetimes and those containing black holes are
qualitatively quite different, there are continuous families of
configurations that connect the two.  In this paper we use
self-gravitating monopole solutions as tools for investigating the
transition between these two types of spacetimes.  We show how
causally distinct regions emerge as the black hole limit is achieved,
even though the measurements made by an external observer vary
continuously.  We find that near-critical solutions have a naturally
defined entropy, despite the absence of a true horizon, and that this
has a clear connection with the Hawking-Bekenstein entropy.  We find
that certain classes of near-critical solutions display naked black
hole behavior, although they are not truly black holes at all.
Finally, we present a numerical simulation illustrating how an
incident pulse of matter can induce the dynamical collapse of a
monopole into an extremal black hole.  We discuss the implications of
this process for the third law of black hole thermodynamics.

\end{abstract}

\setcounter{page}{0}
\thispagestyle{empty}
\maketitle

\eject

\vfill

\baselineskip 16pt plus 2pt minus 2pt

\section{Introduction}

Nonsingular spacetimes and those containing black holes are
qualitatively quite different.  Nevertheless, it is possible to find
sequences of spacetimes that, while remaining nonsingular, come
arbitrarily close to having horizons \cite{orig1,ortiz,orig2,orig2b}.
In a previous paper \cite{LW} we studied a class of such solutions
that are associated with self-gravitating monopoles in a spontaneously
broken Yang-Mills theory.  The emphasis there was on the detailed
behavior of the fields as one approaches the critical solution in
which a horizon first appeared.  In this paper, we concentrate instead
on the geometrical aspects of the spacetimes associated with these
objects near criticality, and on using these to gain insights into the
properties of true black holes.

As in Ref.~\cite{LW}, we restrict ourselves to spherically symmetric
spacetimes and write the metric in the form
\begin{equation}
     ds^2 = B dt^2 - A dr^2 - r^2 (d\theta^2 + \sin^2 \theta\,
     d\phi^2)\ .
\label{metric}
\end{equation}
A horizon corresponds to a zero of $1/A$; the horizon is extremal if
$d(1/A)/dr$ also vanishes.  We work in the context of an SU(2) gauge
theory with gauge coupling $e$ and a triplet Higgs field $\phi$ whose
vacuum expectation value $v$ breaks the symmetry down to U(1).  In
flat spacetime this theory possesses a finite energy monopole solution
with magnetic charge $Q_M=4\pi/e$ and mass $M\sim v/e$.  It has a core
region, of radius $\sim 1/ev$, with nontrivial Higgs and massive
vector boson fields.  Beyond this core is a Coulomb region in which
all massive fields approach their vacuum values exponentially rapidly,
leaving only the Coulomb magnetic field.  The effects of adding
gravitational interactions depend on the value of $v$.  For $v$ much
less than the Planck mass $M_{\rm Pl}$,
one finds that $1/A$ is equal to unity at the origin,
decreases to a minimum at a radius of order $1/ev$, and then increases
again with $A(\infty) =1$.  As $v$ is increased, this minimum becomes
deeper, until an extremal horizon develops at a critical value $v_{\rm
cr}$ of the order $M_{\rm Pl}$.  As we describe in more detail
in Sec.~II, two distinct types of critical behavior are possible,
depending on the ratio of the Higgs and gauge boson masses.  For lower
values of this ratio, one finds ``Coulomb-type'' critical solutions,
in which the horizon occurs in the Coulomb region of the monopole at
$r_0 = \sqrt{4\pi G/e^2}$.  Outside the horizon, the metric is that of
an extremal Reissner-Nordstrom black hole, with
\begin{equation}
      B(r) ={1\over A} = 1 -{2MG\over r} + {Q^2 G\over 4\pi r^2}\ ,
\end{equation}
while the massive fields take on their vacuum values.  As the Higgs
self-coupling increases, there is a transition to ``core-type''
critical solutions that have a horizon inside the monopole core and
nontrivial matter fields (or hair) outside the horizon.  

In both types of critical monopole solutions the fields remain
nonsingular at $r=0$.  However, it is also possible to have solutions
with singularities at $r=0$ that can be viewed as self-gravitating
monopoles with Schwarzschild black holes at their center.  As long as
the mass of the central black hole is not too great, the variation  of
these solutions with $v$ is quite similar to that of the nonsingular
monopoles, and one finds the same two types of critical behavior \cite{BKH}.

After reviewing these solutions, we discuss in Sec.~III how near-critical
monopoles might appear to an outside observer.  One would
expect the measurements made by such an observer to vary continuously
with the parameters of the monopole and to show no discontinuity at
the critical solution.  The external observer could probe the monopole
with either particles or waves.  In the case of the particle, we find that
the time needed for the particle to emerge from the interior (as
measured by a static external observer using ``Schwarzschild time'')
diverges as the critical solution is approached.  When a wave is sent
in, there is a reflected wave due to the gravitational field just
outside the horizon and a transmitted wave that passes through the
interior and then emerges with a time delay.  As before, the time
delay diverges as $v \rightarrow v_{\rm cr}$, while the reflected wave
becomes indistinguishable from that due to a black hole.  Using either
type of probe, an observer whose lifetime is finite cannot distinguish
between a true black hole and a nonsingular, subcritical solution that
is sufficiently close to being critical.  We discuss the implications
for our understanding of black hole entropy.

We also find that the near-critical Coulomb-type solutions display
what Horowitz and Ross \cite{naked1,naked2} have termed
naked-black-hole behavior, even though there is no black hole at all.
This is characterized by the fact that a freely-falling observer
passing through the minimum of $1/A$ (we shall refer to the location
of this minimum as the quasi-horizon) feels a tidal force that
diverges as the critical solution is approached.  For core-type
solutions, on the other hand, no such behavior is observed.

In Sec.~IV, we consider the effect of having additional matter fall
into a near-critical solution, addressing in particular the question
of whether this process could produce an extremal black hole.
Extremal black holes are especially interesting from the standpoint of
black hole thermodynamics because they have vanishing Hawking
temperature.  The analogies between black hole dynamics and
thermodynamics thus suggest that they should be rather difficult, if
not impossible, to create.  Indeed, one of the formulations
\cite{third} of the third law of black hole dynamics asserts the
impossibility (under certain technical assumptions) of making a
nonextremal black hole extremal.  One could also envision producing an
extremal black hole starting from a nonsingular spacetime.  Boulware
\cite{boulware} showed that this can be done by
the collapse of a charged shell of matter.  However, this mechanism
relies critically on the shell being infinitely thin; shells of finite
thickness and density do not collapse to an extremal configuration.

It is easy to understand the difficulty of making an extremal black
hole if one recalls that the extremal Reissner-Nordstrom black hole is
characterized by having a mass and a charge that (in appropriately
rescaled Planck units) are equal.  Forming such an object by the
collapse of a shell with equal charge and mass densities involves a
delicate balance between electromagnetic and gravitational forces.
One could instead try to achieve extremality by adding matter to a
pre-existing nonextremal Reissner-Nordstrom black hole (i.e. one with
greater mass than charge).  However, because the added matter would
have to have more charge than mass, the Coulomb repulsion between the
black hole and the infalling matter would tend to overcome their
gravitational attraction.

The situation is rather different in our case, because the nonsingular
monopole solutions are {\it overcharged}; i.e., their long range
fields are those of a Reissner-Nordstrom solution with greater charge
than mass.\footnote{In the pure Reissner-Nordstrom case, this leads to
a naked singularity.  The singularity is avoided here by the same
mechanism that makes the mass of the flat-space monopole finite: the
orientation of the massive gauge fields in the monopole core is such
that their magnetic dipole energy just cancels the singular Coulomb
energy at the origin.}  Allowing uncharged matter to fall into these
objects increases their mass and should bring them closer to
criticality.  If the amount of matter entering is just sufficient to
create a zero of $1/A$, one would expect an extremal solution to
result.  We will present numerical arguments that support this
expectation.  Finally, we make some concluding remarks in Sec.~V.

\section{Review of previous results}

As in Ref.~\cite{LW}, we consider an SU(2) gauge theory that is
spontaneously broken to U(1) by the vacuum expectation value $v$ of a
triplet Higgs field $\phi$.  This theory has magnetic monopole
solutions that can be described by the spherically symmetric ansatz
\begin{eqnarray}
\phi^a &=& v {\hat r}^a h(r)
\label{phiansatz}	\\
A_{ia} &=& \epsilon_{iak} {\hat r}^k\, {1 -u(r) \over er}\ .
\label{Aansatz}		
\end{eqnarray}
Finiteness of the energy requires that $u(\infty) = 0$ and 
$h(\infty) = 1$.  If the solutions are also required to be nonsingular at
$r=0$, then $u(0)=1$ and $h(0)=0$.

In a spacetime with a metric of the form of Eq.~(\ref{metric}), the
static field equations for these matter fields can be derived from
a $(1+1)$-dimensional action of the following form \cite{vanN}
\begin{equation} 
   S_{\rm matter} = -4\pi \int dt\, dr\, r^2 \sqrt{AB} 
      \left[ {K(u,h)\over A} + U(u,h) \right]
\label{matteraction}
\end{equation}
where $U(u,h)$, involves the fields but not their derivatives,
while
\begin{equation} 
   K = \frac{1}{e^2r^2}\left(\frac{du}{dr}\right)^2 
	+ \frac{v^2}{2}\left(\frac{dh}{dr}\right)^2\ .
\end{equation}

The Euler-Lagrange equations for the matter fields that follow from
Eq.~(\ref{matteraction}) must be supplemented by the gravitational
field equations.  For static, spherically symmetric field configurations
these reduce to 
\begin{equation} 
    G_{\hat t\hat t} = {1\over 2 r^2} {d\over dr}\left[r\left({1\over
    A} -1\right) \right] = - 4\pi G \left({K\over A} + U \right)
\end{equation}
and 
\begin{equation} 
   G_{\hat t\hat t}+ G_{\hat r\hat r} = -\left({2 \over rA}\right)
     \frac{1}{\sqrt{AB}}\frac{d\sqrt{AB}}{dr} = -{16\pi G  K \over A}\ .
\label{ABeqn}
\end{equation}
Here carets indicate orthonormal components.

Equation~(\ref{ABeqn}) can be immediately integrated to obtain 
\begin{equation}
   B(r) = {1 \over A(r)} \exp\left[-16\pi G\int_r^\infty dr'\, r'
   K\right]\ .
\label{Bsolution}
\end{equation}
Using this to eliminate $B$, one is left with two second order and one
first order equation for the functions $u$, $h$, and $A$.  These
equations must be solved numerically.  Up to a rescaling of distances,
the solutions of these equations depend only on the two dimensionless
parameters ${a} = 8\pi Gv^2$ and ${b} = (m_H/2 m_W)^2$.

For small values of $b$ (roughly $b \lesssim 25$ for a quartic Higgs)
\cite{LW,BKH} one finds Coulomb-type solutions in which the minimum of
$1/A$ is located outside the monopole core.  This minimum
decreases\footnote{This behavior is modified slightly for very small
$b$.  For a detailed description, see
\cite{ortiz,orig2,stability,stability2}.} with increasing $a$, until
the critical solution is reached at $a_{\rm cr} =8\pi Gv^2_{\rm cr}$.
In the critical solution, the matter fields $u$ and $h$ reach their
asymptotic values $u=0$ and $h=1$ at the horizon and are then constant
for all $r >r_0$; because both fields fall as fractional powers of
$r_0 - r$, the derivatives $du/dr$ and $dh/dr$ both diverge as $r$
approaches $r_0$ from below.  (This nonanalytic behavior is possible
because an extremal horizon is a singular point of the matter field
equations.)

The metric of the critical solution is identical to the extremal
Reissner-Nordstrom 
metric outside the horizon, but differs from it for $r < r_0$.  The
metric function $1/A$ varies relatively smoothly, falling
monotonically from unity at the origin to a zero at the horizon.  Just
inside the horizon $1/A \sim k (r_0 -r)^2$, with $k$ being larger than
for the corresponding Reissner-Nordstrom solution. The behavior of $B$
contrasts sharply with 
this.  Equation~(\ref{Bsolution}) shows that the product
$AB$ (which is identically equal to unity in both the Schwarzschild
and Reissner-Nordstrom solutions) is given by an integral of the
functional $K(u,h)$.  The singularities in the derivatives of $u$ and
$h$ at the horizon are strong enough to cause this integral to
diverge, so that the ratio
\begin{equation}
       c \equiv  {\sqrt{AB}|_{{\rm outside}\ r_0} \over
             \sqrt{AB}|_{{\rm inside}\ r_0} }
\end{equation}
is infinite in the critical limit.  If we adopt the conventional
normalization $B(\infty)=1$, then $B$ vanishes identically inside the
horizon.  If we instead set $B(0)=1$, then $B$ is finite and varying
inside the horizon and infinite for $r>r_0$; depending on the value of
$b$, the minimum of $B$ may be at $r=0$ or at some finite radius, but
in neither case does $B$ have a zero.  For near-critical solutions
where the minimum value $(1/A)_{\rm min} \equiv \e$ is small but
nonzero, we find that the ratio $c$ varies as $\e^{-q}$, where
$q$ ranges from about 0.7 to unity.

\begin{figure} \begin{center}\PSbox{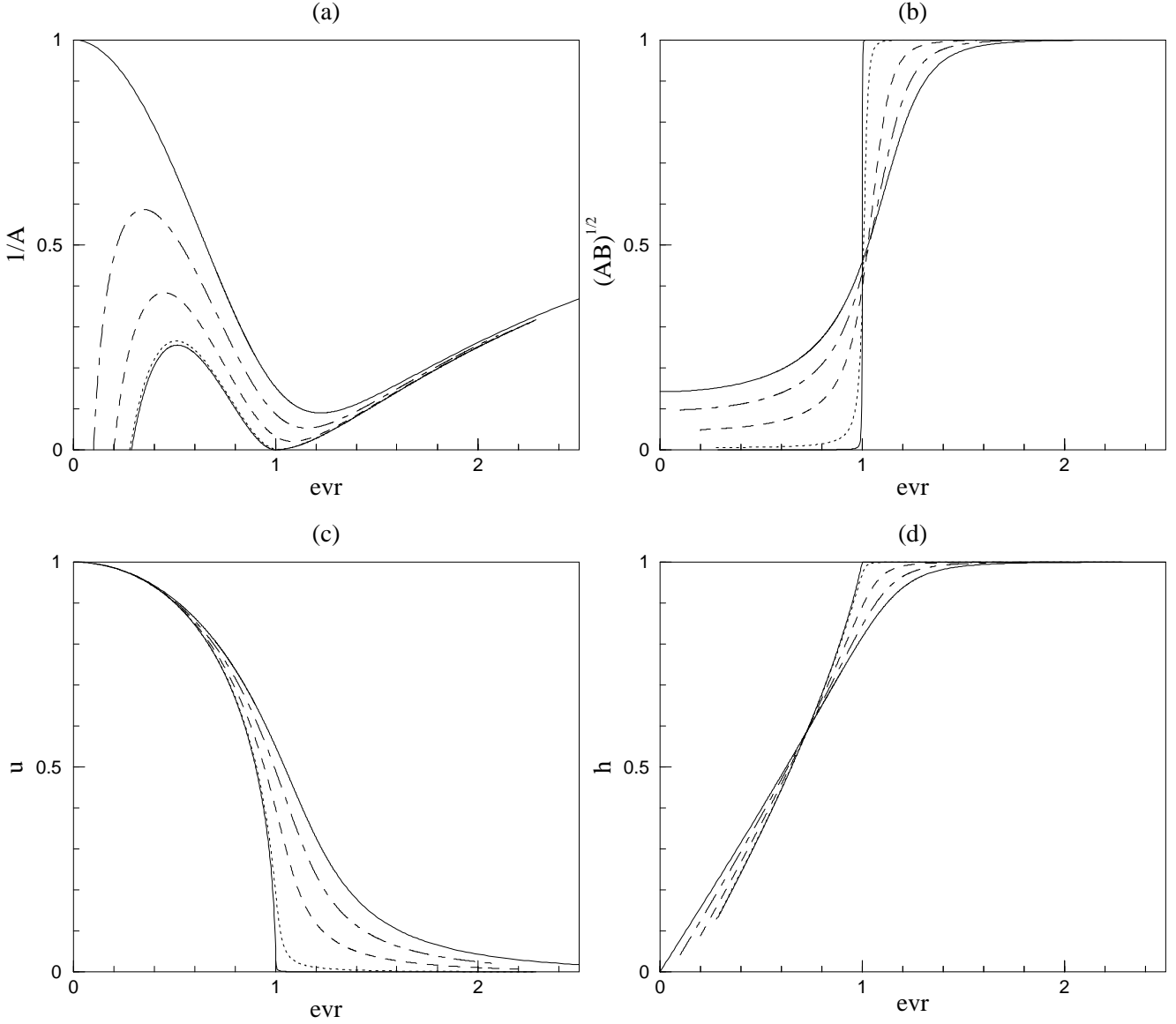
hscale=100 vscale=100 hoffset=-120 voffset=-35}{5in}{6in}\end{center}
\caption{
Monopole solutions for $a=2.0$, ${{b}} = 1.0$ and various
values of central black hole radius, $r_H$.  The progression from
solid line, dot-dashed line, to dashed line, to dotted line, to solid
line corresponds to ${{evr_H}} = 0.0, 0.1, 0.2, 0.28$ and $0.288$.
The panels depict the functions (a) $1/A(r)$, (b) $(AB)^{1/2}(r)$, (c)
$u(r)$ and (d) $h(r)$.
}
\label{fig:low}
\end{figure}

\begin{figure} \begin{center}\PSbox{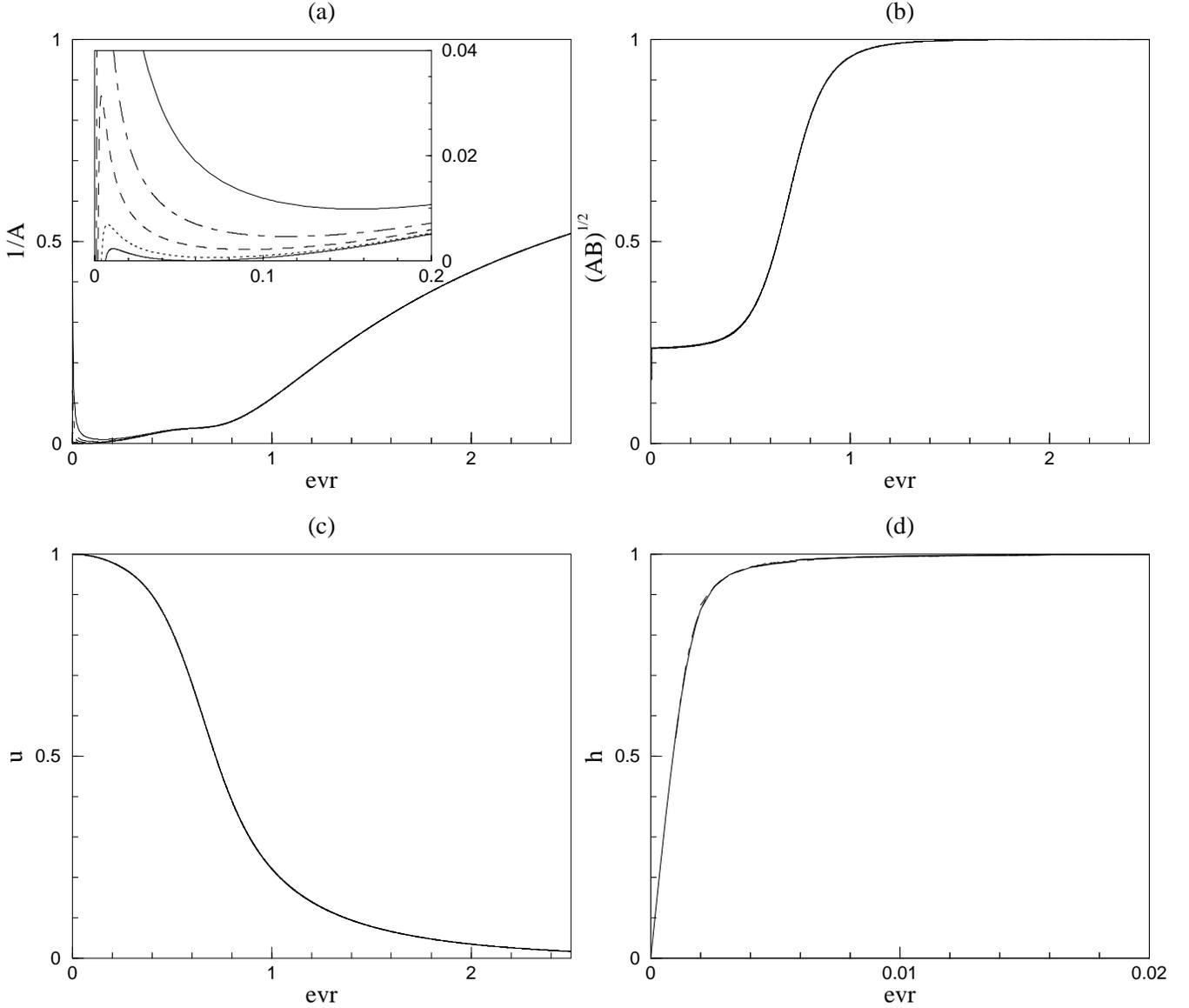
hscale=100 vscale=100 hoffset=-120 voffset=-35}{5in}{6in}\end{center}
\caption{
Monopole solutions for $a=1.002$ and ${{b}} = 10^6$ and various values
of central black hole radius, $r_H$.  Here $a_{\rm cr}(r_H = 0) =
1.011654$ and the minimum $a$ using this scenario is $1.001$.  The
progression from solid line, dot-dashed line, to dashed line, to
dotted line, to solid line corresponds to ${{evr_H}} = 0, 0.001,
0.002, 0.004$ and $0.00628$.  The panels depict the functions (a)
$1/A(r)$, (b) $(AB)^{1/2}(r)$, (c) $u(r)$ and (d) $h(r)$.  Note that
the monopole fields are virtually unchanged as the internal black hole
size is varied.  Note that the radial scale for $h(r)$ is exaggerated to
show detail.
}
\label{fig:big}
\end{figure}

A rather different type of critical solution is found for larger $b$.
For these core-type solutions the horizon occurs at a radius $r_* <
r_0$, with the values $u_*$ and $h_*$ of the matter fields at this
point being different than their asymptotic values.  Although the
solutions are still nonanalytic at the horizon, this nonanalyticity
occurs only in subdominant terms.  Thus, $1/A$ again vanishes as
$(r-r_0)^2$ as one approaches the horizon, but the coefficient is now
the same inside and outside the horizon.\footnote{For intermediate
values of $b$, subcritical monopoles exhibit both core-type and
Coulomb-type quasi-horizons.  However, as one approaches criticality
for a given value of $b$, only one quasi-horizon actually becomes a
horizon.  The other quasi-horizon, though interesting, is essentially
irrelevant for our purposes.} The radial derivatives of both matter
fields are finite at the horizon, so $K$ remains finite and there is
no sharp change in $AB$ at the horizon.  Because $AB$ remains finite
and nonzero, $B$ has a zero at the horizon that coincides with the
zero of $1/A$.

These solutions can be generalized to include a black hole in the
center of the monopole.  Instead of requiring that the fields be
nonsingular at $r=0$, one instead requires that there be a zero of
$1/A$ at a nonzero radius $r_{\rm H}$.  At this zero, the equations
for the matter fields become constraint equations relating the fields
and their first derivative; solving these constraints yields enough
boundary conditions to determine a solution.

If $r_{\rm H}$ is not too large, the effect of increasing $a$ is similar
to what it is in the absence of a central black hole.\footnote{
For larger values of $r_H$, see the discussion in \cite{orig2}.}
There is an outer minimum of $1/A$ that moves
downward, finally reaching zero and becoming an extremal horizon at
some critical value $a_{\rm cr}(r_{\rm H})$.  For small values of $b$
the solutions are Coulomb-type, while for large $b$ one finds
core-type critical solutions.

Rather than increasing $a$ with $r_{\rm H}$ fixed, one can instead
increase $r_{\rm H}$ with $a$ held fixed; this is much more analogous
to the process of actually dropping matter into a near-critical
solution that we will consider in Sec.~IV.  For initial values of $a$
that are sufficiently close to $a_{\rm cr}(r_{\rm H}=0)$, this gives a
family of solutions with a critical limit.  In Figs.~\ref{fig:low} and
\ref{fig:big} we illustrate this with Coulomb-type solutions with
$b=1.0$ and core-type solution with $b=10^6$.

\section{Probing the almost black hole}

\begin{figure} \begin{center}\PSbox{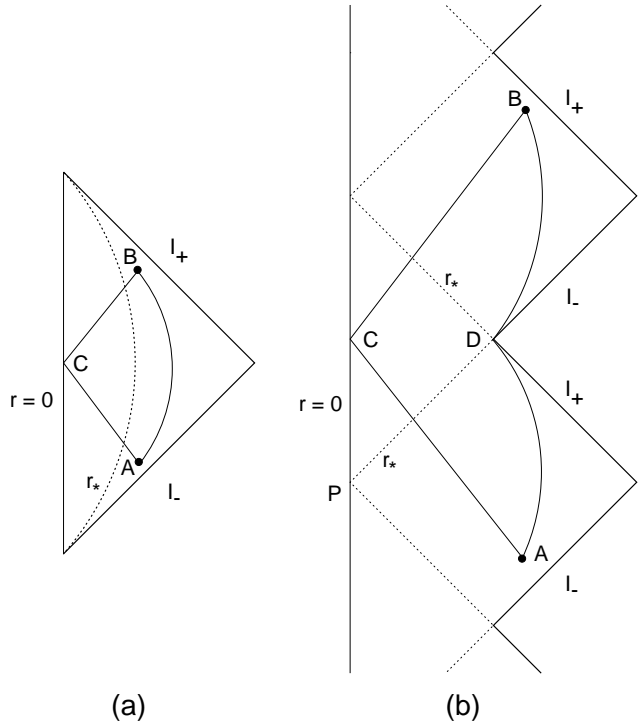
hscale=50 vscale=50 hoffset=50 voffset=0}{4in}{3.5in}\end{center}
\caption{
Penrose diagrams for (a) subcritical monopole and (b) critical
monopole black hole.  In the former case $r_*$ represents the
quasi-horizon whereas in the latter case, that radius represents
a true horizon.
}
\label{fig:penrose}
\end{figure}

For any $a<a_{\rm cr}$, the self-gravitating monopole solution is
a nonsingular spacetime with a Penrose diagram of the same form as
that of Minkowski spacetime (Fig.~\ref{fig:penrose}a).  The critical
solution, on the other hand, can be extended beyond the original
coordinate patch to yield a spacetime with the Penrose diagram shown
in Fig.~\ref{fig:penrose}b.  This diagram is quite similar to that of
an extremal Reissner-Nordstrom black hole, but differs from it by not
having a singularity\footnote{Although there is no singularity at the
origin for a critical monopole black hole, there are singularities at
the extremal horizon resulting from nonanalytic behavior of the
monopole fields.  These singularities are relatively mild in the
core-type case, but are more dramatic in the Coulomb-type case.} at
$r=0$.  The difference between the two diagrams is striking and seems
to indicate a discontinuity at $a=a_{\rm cr}$, in contradiction with
the usual expectation that physical quantities should vary
continuously with the parameters of a theory.

However, this discontinuity can been seen as an artifact of the
conformal transformation that produces the Penrose diagram from an
infinite spacetime.  This can be illustrated by considering the points
A and B that lie on the curves of constant $r$ that are shown in
Figs.~\ref{fig:penrose}a and \ref{fig:penrose}b.  These have been chosen so
that it is possible for an object to start at A, move in to $r=0$ at
C, and then travel out again to B.  The total proper time along this
worldline (or the total affine parameter, if the worldline is
lightlike) is finite.  This should be compared with the proper time
along the worldline of constant $r$.  This is finite for the
subcritical case, whereas in the critical case the proper times along
the segments AD and DB are both infinite, corresponding to the fact
that an observer in the exterior region containing A cannot detect
objects behind the horizon PD.  In order to obtain consistency with
our physical expectations of continuity, we should require that the
proper time along the world of constant $r$ should diverge as $a
\rightarrow a_{\rm cr}$.  More generally, the time required for an
external observer to receive information from a probe of the interior
regions solution should diverge in the critical limit.

\subsection{Particle and Wave Probes}

To see how this works out, we consider probing the interior region
(i.e., the region $r < r_*$, where $r_*$ is the quasi-horizon) of a
near-critical solution by sending in either a particle or a wave.  In
both cases, we assume that the probe interacts only gravitationally,
and has no direct interaction with the monopole fields.

To begin, imagine releasing a massive particle from an initial radius
$r_1 \gg r_*$ that is large enough that we may approximate $B(r_1)
\approx 1$.   The rotational and time-translation symmetries of the
metric allow us to take the motion to lie in the $\theta=\pi/2$ plane
and guarantee the conservation of the angular momentum per unit mass
\begin{equation}
     J = r^2 {d\phi \over d\tau}
\label{J}
\end{equation}
and the energy per unit mass
\begin{equation}
     E = B(r) {dt \over d\tau}\ .
\end{equation}
These, together with Eq.~(\ref{metric}), imply that
\begin{equation}
     {dr \over d\tau}  = {1 \over \sqrt{AB}} 
           \left[ E^2 - B\left({J^2 \over r^2}
      + 1  \right) \right]^{1/2}
\label{drdt}
\end{equation}
If $J=0$, the particle falls radially in, pass through the origin,
and emerge on the other side of the monopole.  If instead $J\ne 0$, the
particle turns around after reaching a minimum radius $r_{\rm
min}(J)$ and return to $r_1$ with its trajectory advanced by an angle
\begin{equation}
    \Delta \phi = 2 \int_{r_{\rm min}}^{r_1} dr\, {d\phi/d\tau \over
                   dr/d\tau} 
       = 2 \int_{r_{\rm min}}^{r_1} dr \,
      {J  \over r^2}\sqrt{AB} \left[ 1 -{B \over E^2} \left({J^2 \over r^2}
      + 1 \right) \right]^{-1/2}\ .
\end{equation}
We are interested in the proper time measured by an observer who
remains at $r=r_1$.  Assuming that the observer is not moving at
relativistic speeds, this is approximately the same as the
Schwarzschild coordinate time $t$, which over the course of the entire
trajectory increases by an amount  
\begin{equation}
    \Delta t = 2 \int_{r_{\rm min}}^{r_1} dr\, {dt/d\tau \over
                   dr/d\tau} 
       = 2 \int_{r_{\rm min}}^{r_1} dr\,
     {A \over \sqrt{AB}} \left[ 1 -{B \over E^2} \left({J^2 \over r^2}
      + 1\right) \right]^{-1/2}\ .
\label{dt}
\end{equation}
There are two potential sources of divergences in this integral as
$\e =(1/A)_{\rm min} \rightarrow 0$ and the critical solution is
approached.  In both types of critical solutions there is a
contribution from $r \approx r_*$ associated with the growth of
$A(r_*)$.  In the Coulomb-type solutions the near-vanishing of
$\sqrt{AB}$ [see Eq.~(\ref{Bsolution})] gives a second contribution from
the entire region $r<r_*$.  Let us examine these in more detail.

For core-type solutions, in the region $r \approx r_*$ we can write
\begin{equation}
    A \approx  k_1\left[ \left(r-r_*\over r_*\right)^2 +  \e\right]^{-1}
\end{equation}
with $k_1$ of order unity, while $\sqrt{AB}$ is roughly constant and
independent of $\e$.  Because $B(r_*)$ is small, the $J$-dependent
term in $\Delta t$ can be neglected for any $J$ such that the particle
could have reached $r_*$.  By a similar argument, we see that any
particle that reaches $r_*$ goes through the peak of $A$ before
turning around. Hence,
\begin{equation}
        \Delta t \approx {2 k_1 \pi r_* \over \sqrt{AB}|_{r=r_*}}
               \e^{-1/2}   + \cdots
\end{equation}
where the ellipses represent subdominant terms.

For Coulomb solutions the dominant effect is due to the fact that
$\sqrt{AB} \sim \e^q$ is almost vanishing throughout the
interior region.  Because our numerical solutions show that $q$ ranges
between 0.7 and unity, the divergence due to this effect is greater
than that from the region near $r_*$.  Furthermore, the near-vanishing
of $B$ in the interior implies that any particle that enters the
interior almost reaches the origin, so that $r_{\rm min} \approx 0$.
Thus,
\begin{equation}
        \Delta t \approx k_2 r_*   \e^{-q}
               + \cdots
\end{equation}
where $k_2$ is of order unity.

Rather than sending in a particle, one can also probe the near-black
hole by sending in a wave packet.  As an example, let us consider a free
massive scalar field $\phi$, whose field equation in curved spacetime
takes the form
\begin{equation}
     0 = {1\over \sqrt{g}} \partial_\mu\left[\sqrt{g} g^{\mu\nu}
     \partial^\nu \phi \right] + m^2 \phi\ .
\label{scalar}
\end{equation}
This can be put into a more tractable form by writing
\begin{equation}
      \psi = r \phi
\end{equation}
and defining a new coordinate $y(r)$ satisfying
\begin{equation}
    {dr \over dy} = {\sqrt{AB} \over A}\ .
\label{newcoord}
\end{equation}
Equation~(\ref{scalar}) then takes the form of a one-dimensional wave
equation
\begin{equation}
     0 =  {\partial^2 \psi \over \partial t^2} 
     - {\partial^2 \psi \over \partial y^2}  + \left [U(r) + m^2
          B\right]  \psi
\end{equation}
with a scattering potential
\begin{equation}
   U(r) = { 1\over 2r} {d \over dr}\left[{AB \over
          A^2}\right]  +  {J(J+1) B\over r^2}\ .
\end{equation}

When a wave packet incident from large $r$ reaches the region near the
quasi-horizon, a portion is reflected by the scattering potential,
while a portion is transmitted and emerge with some time delay.
If a near-critical solution is to appear essentially
indistinguishable from a black hole to an outside observer, two
conditions must hold.  First, the reflection coefficient as a function
of wave number must approach that of the black hole as $\e
\rightarrow 0$.  Second, the time delay in the emergence of the
transmitted wave should diverge in the critical limit.   

To see how the first of these conditions comes about, let us use
Eq.~(\ref{ABeqn}) to rewrite the scattering potential as
\begin{equation}
   U(r) = { AB\over rA} \left[{8 \pi G K\over A} - {d\over
          dr}\left({1\over A} \right)  +  {J(J+1)\over r} \right]\ .
\end{equation}
For both core- and Coulomb-type solutions the quantity $K/A$ remains
finite in the critical limit, while the
second term in the brackets is zero at the quasi-horizon.  Since $AB
\le 1$, it is clear that $U(r_*)$ vanishes at least as fast as
$\e$ as the critical limit is approached.   Hence, in the limit
the scattering potential splits into two parts, one inside and one
outside the horizon.  The outer potential is equal either to that of an
extremal Reissner-Nordstrom black hole (in the Coulomb case) or that
of a black hole with hair (in the core case).  Because the variation
of the outer potential with $\e$ is smooth in both cases, our
conditions on the reflection coefficients are satisfied if we can
ignore reflection from the inner part of the potential.

This can be understood by noting that the natural distance variable in
which to discuss the motion of the waves is $y$.  By integrating
Eq.~(\ref{newcoord}) inward from some reference point $r_1 \gg r_*$,
we obtain
\begin{equation}
      y(r) = y(r_1) - \int_r^{r_1} dr \, {A \over \sqrt{AB}}\ .
\end{equation}
The behavior of this integral as the critical limit is approached is
very similar to that of the integral in the expression for $\Delta t$,
Eq.~(\ref{dt}).  For either type of solution, the region near the
quasi-horizon gives a contribution that diverges at least as fast as
$\e^{-1/2}$.  There is a corresponding growth in both the
effective distance from the inner portion of the potential to any
external point and in the time delay of the corresponding reflected
wave.  As $\e$ is increased, an external observer at fixed $r$
first sees the reflections from the inner and outer parts of the
potential split into two distinct reflected waves, and then finds
that the time delay of the second reflected wave (from the inner
potential) grows without bound.  

The portion of the wave that is transmitted through the region near
the quasi-horizon either continues through the origin and then outward or
reflects off a central centripetal barrier, according to
whether or not $J$ vanishes.  In either case, the time delay
accumulated by this wave before it returns to the quasi-horizon
grows in essentially the same manner as the travel time for a massive
particle traversing the same path: as $\e^{-1/2}$ for a
core-solution and as $\e^{-q}$ for a Coulomb-type solution.

\subsection{Information and Entropy}

Thus, regardless of the type of probe used, an external observer at
fixed $r_{\rm obs}$ must wait for at least a time $\Delta t \ge
O(\e^{-1/2})$ before the probe emerges from the region inside the
quasi-horizon.  To leading order, this time delay is independent of the
energy or angular momentum of the probe, and is instead determined solely
by the spacetime geometry.  Hence, to an observer with a finite
lifetime $T$, the interior region of any near-critical configuration
with $\e \lesssim T^{-2}$ is inaccessible.\footnote{Note that once
$\e$ is less than $T^{-2}$, the the boundary of the inaccessible
region depends only very weakly on $T$ and $r_{\rm obs}$, and is
essentially indistinguishable from the quasi-horizon.} He would most
naturally describe any larger system containing this configuration by
a density matrix $\rho$ obtained by tracing over the degrees of
freedom inside the quasi-horizon.  From this density matrix one can
derive an entropy $S_{\rm interior}= - {\rm Tr}\,\rho \ln \rho$ that
can be associated with the interior of the quasi-black hole.

One could, of course, proceed in this manner to define an entropy for
any arbitrary region in space, just as one can choose to make the
information in any subsystem inaccessible by putting the subsystem
behind a locked door.  The crucial difference here is that the
inaccessibility is due to the intrinsic properties of the spacetime,
and that the boundary of the inaccessible region is defined by the system
itself rather than by some arbitrary external choice.  It is thus
reasonable to define $S_{\rm interior}$ as {\it the} entropy of the
quasi-black hole.

A precise calculation of this entropy is clearly infeasible.  Among
other problems, such a calculation would require a correct
implementation of an ultraviolet cutoff, which presumably would
require a detailed understanding of how to perform the calculation in
the context of a consistent theory of quantum gravity.  As an initial
effort, one can take the ultraviolet cutoff as the Planck mass $M_{\rm
Pl}$ and ask for an order of magnitude calculation.  Such a
calculation was done by Srednicki \cite{srednicki}, who showed that
tracing over the degrees of freedom of a scalar field inside a region
of flat spacetime with surface area $A$ led to an entropy $S=\kappa
M^2 A$ where $M$ is the ultraviolet cutoff and $\kappa$ is a numerical
constant.  Furthermore, although the precise calculations depend on
the details of the theory, Srednicki gave general arguments
suggesting that such an entropy should always be proportional to the
surface area.  This leads us to expect that $S_{\rm interior} \sim
M_{\rm Pl}^2 A$.

This result is, of course, consistent with the possibility that in the
critical limit $S_{\rm interior}$ goes precisely to the standard black
hole result $S_{\rm BH} = M_{\rm Pl}^2 A/4$.  However, in contrast
with the usual black hole case, our spacetime configurations are
topologically trivial.  The ``interior'' region enclosed by the
quasi-horizon is nonsingular and static.  Furthermore, this region can
be unambiguously defined, so that it is conceptually clear what is
meant by tracing over the interior degrees of freedom, even though it
may not yet be possible to implement this calculation in complete
detail.  We find it quite striking that by this approach one can
arrive so nearly at the standard entropy result.

\subsection{Curvature and Naked-Black-Hole Behavior}

In our discussion above of the trajectory of a particle probe, we
focussed on the coordinate time that elapses over the course of the
particle's passage through the monopole.  However, it also of interest
to consider the elapsed proper time, which can be found by integrating
$d\tau/dr$ [see Eq.~(\ref{drdt})].  For core-type solutions this gives
a finite nonzero result with no unusual behavior as the critical limit
is approached.  The situation with Coulomb-type solutions is, on the
other hand, quite striking.  The sharp decrease in $AB$ at the
quasi-horizon leads to a corresponding decrease in $d\tau/dr$, so that
the proper time elapsed while the probe is within the quasi-horizon
is\footnote{The drop in $AB$ also has consequences for the shape of
the trajectory through the interior.  By combining Eqs.~(\ref{J}) and
(\ref{drdt}), one finds that in the critical limit all probes follow a
straight line passing through the origin, regardless of the incident
angle with which they hit the horizon.}
\begin{equation}
    \Delta \tau \approx {2 r_* \over E}  \sqrt{AB}|_{r=0} \sim
    \epsilon^q\ .
\end{equation}
In the critical limit $AB$ vanishes identically for $r < r_*$, and
$\Delta \tau =0$.

This vanishing of $\Delta \tau$ is related to another interesting
property of these solutions.  It is well known that the Riemann tensor
is nonsingular at a black hole horizon.  It therefore does not seem
surprising that in the most familiar black hole solutions, the
Schwarzschild and Reissner-Nordstrom, a particle suffers no unusual
effects as it crosses the horizon.  However, Horowitz and Ross
\cite{naked1} showed that this is not always the case.  Because of the
acceleration of a particle as it approaches the horizon, the
components of the Riemann tensor in a coordinate frame that is freely
falling with the particle can be quite different from the components
measured in a static frame.  With a metric of the form of
Eq.~(\ref{metric}), the components $R_{t'kt'k}$ (where $k$ denotes a
transverse spatial direction and $t'$ the time in the boosted frame)
are given by
\begin{equation}
    R_{t'kt'k} = -{1\over 2r} {d\over dr} \left[ {E^2 \over AB} -
    {1\over A} \right]\ ,
\label{nakedbh}
\end{equation}
where $E$ is the energy per unit mass of the infalling particle.

The fact that this curvature component is never large (with $E$ of
order unity) for the Schwarzschild and Reissner-Nordstrom black holes
is a consequence of the fact that $AB$ is constant in both cases.
This is not true in general.  Horowitz and Ross found several examples
of dilaton black holes for which $R_{t'kt'k}$, and thus the tidal
forces felt by an infalling particle, could be made arbitrarily large
near the horizon by taking the solution to be sufficiently close to
extremality.  They introduced the term naked black hole to indicate
the fact that this (almost) singular behavior occurs outside the
horizon.  Subsequently \cite{naked2}, they showed that in these examples
$R_{t'kt'k}$ was inversely proportional to the square of the proper
time remaining before the particle reached the singularity at $r=0$.

Applying their results to our solutions, we find that near-critical
Coulomb-type solutions display naked black hole behavior, even though
they are not black holes at all.  This can be seen by noting that 
Eq.~(\ref{ABeqn}) implies that 
\begin{equation}
    {d\over dr} \left( {1 \over AB}\right) = {16 \pi G r K \over AB}\ .
\end{equation}
Inside the quasi-horizon $AB \sim \epsilon^{2q}$, while the radial
derivatives of $u$ and $h$, and hence $K$, are of order unity.
Inserting this result into Eq.~(\ref{nakedbh}), gives
\begin{equation}
    R_{t'kt'k} \sim \epsilon^{-2q}\ .
\end{equation}
Note that this is proportional to $(\Delta \tau)^{-2}$, giving a
relationship between tidal forces and proper time reminiscent of the
examples described in Ref.~\cite{naked2}.

\section{Collapse to an Extremal Black Hole}

To gain an insight into the third law of thermodynamics, as noted in
Sec.~I, it is of interest to determine whether systems with either
initially nonsingular spacetimes or initially non-extremal black holes
can evolve into systems with an extremal horizon.  It appears that
under reasonable conditions of finiteness and causality this cannot be
done by adding charge to an undercharged object \cite{third,boulware}.
The discussion in the previous two sections, however, suggests an
alternative.  Recall that our subcritical monopoles are overcharged;
i.e., they have a charge larger than their mass.  In the normal
Reissner-Nordstrom case, such a system would exhibit a naked
singularity.  However, in the monopole the Coulomb core is screened by
the massive particles in such a way that no gravitational singularity
exists.  This suggests a scenario by which an extremal
black hole is dynamically formed from a monopole by dropping in
uncharged matter.

Let us add to our theory a chargeless matter field that is coupled to
the monopole fields only through gravity.  We then allow a spherical
shell (of small, but finite thickness) of this matter fall into the
monopole.  If the mass of the shell is sufficiently small, we do not
expect a horizon to be formed.  On the other hand, if the shell
contains enough matter, the system should collapse to form a black
hole.  It seems plausible that threshold case between these two
regimes should produce an extremal horizon.

In the case of a Coulomb-type solution, one might run into
difficulties because of the naked-black-hole behavior it exhibits.
This concern, however, should not be an issue for the case of a
core-type solution.  Here the fields of the critical solution are much
better behaved; what nonanalyticities exist at the horizon are mild,
and become increasingly so as one increases $b$.  Moreover, nothing
unusual happens to the near-critical core-solutions as one
parametrically approaches criticality.  Adding a small Schwarzschild
black hole at the center of the self-gravitating monopole should not
significantly change the scenario.  The infall of a spherical shell of
appropriate mass should still turn the quasi-horizon of a
near-critical solution into an extremal horizon.  The black hole would
essentially play a spectator role, as its presence is largely
irrelevant to the dynamics of the system.

We have carried out numerical simulations to test these ideas.  We
begin with a core-type monopole solution that, for numerical
convenience, has a small Schwarzschild black hole with horizon radius
$r_H$ at its center.  The parameters are chosen to be such that the
solution is near criticality, so that only a small amount of
additional matter is needed for the quasi-horizon at $r=r_*$ to
collapse to a true horizon.  We add to the theory a massive scalar
field $\chi(r,t)$ that is coupled only to gravity.  We then send a
spherically symmetric Gaussian pulse of $\chi$ field in toward the
monopole and watch the system evolve.  To simplify the computation, we
freeze the matter field variables $u$ and $h$ at their initial values;
because the fields for near-critical core-type monopoles are not very
sensitive to the metric (cf. Fig~\ref{fig:big}), this approximation
should cause little error.

When the pulse amplitude is larger than some threshold value, the
pulse falls into the monopole until the metric function $1/A$ develops
a simple zero near the location of the quasi-horizon of the initial monopole
configuration.  This newly formed horizon is non-extremal.  More
interesting is the situation where the pulse amplitude is at, or just
below, this threshold value.  Figure~\ref{fig:t1} shows a sequence of
snapshots illustrating this scenario at four different points during
the pulse's motion inward.\footnote{Note that we have chosen a
normalization of time such that $AB \rightarrow 1$ as $r \rightarrow
r_H$, the horizon of the internal black hole; this corresponds to
using a time variable appropriate to an observer in the interior of
the monopole.  If we had used the more conventional normalization with
$AB$ equal to unity at spatial infinity, the time coordinate would be
that appropriate to an external observer and would grow rapidly as the
pulse approached the quasi-horizon.}  (The distortion of the pulse is
due to the backreaction of the monopole metric.)

\setcounter{figure}{3}
\begin{figure} \begin{center}\PSbox{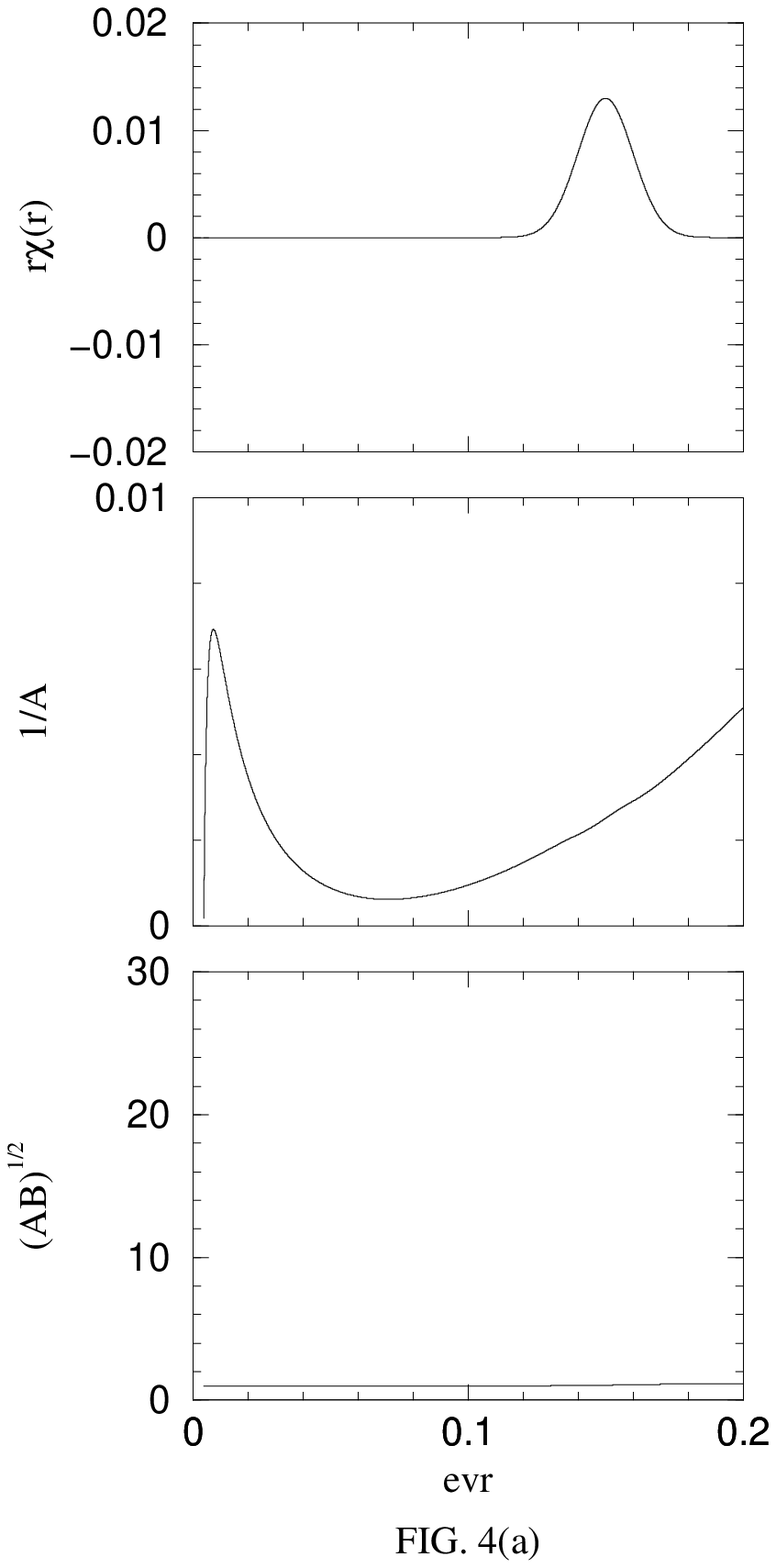
hscale=100 vscale=100 hoffset=-200 voffset=-15}{3in}{6in}\end{center}
\label{fig:t1}
\end{figure}

\setcounter{figure}{3}
\begin{figure} \begin{center}\PSbox{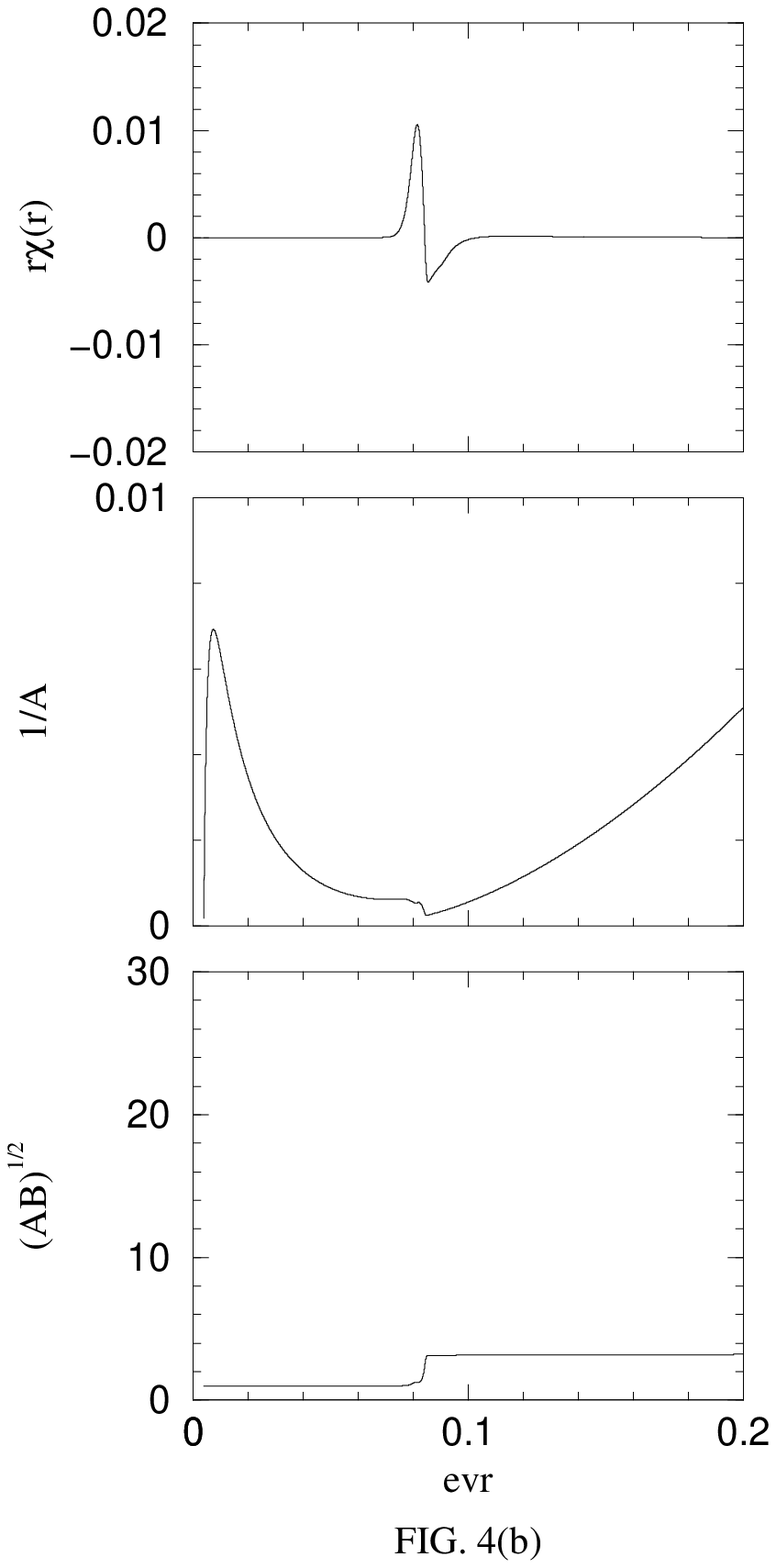
hscale=100 vscale=100 hoffset=-200 voffset=-15}{3in}{6in}\end{center}
\label{fig:t2}
\end{figure}

\setcounter{figure}{3}
\begin{figure} \begin{center}\PSbox{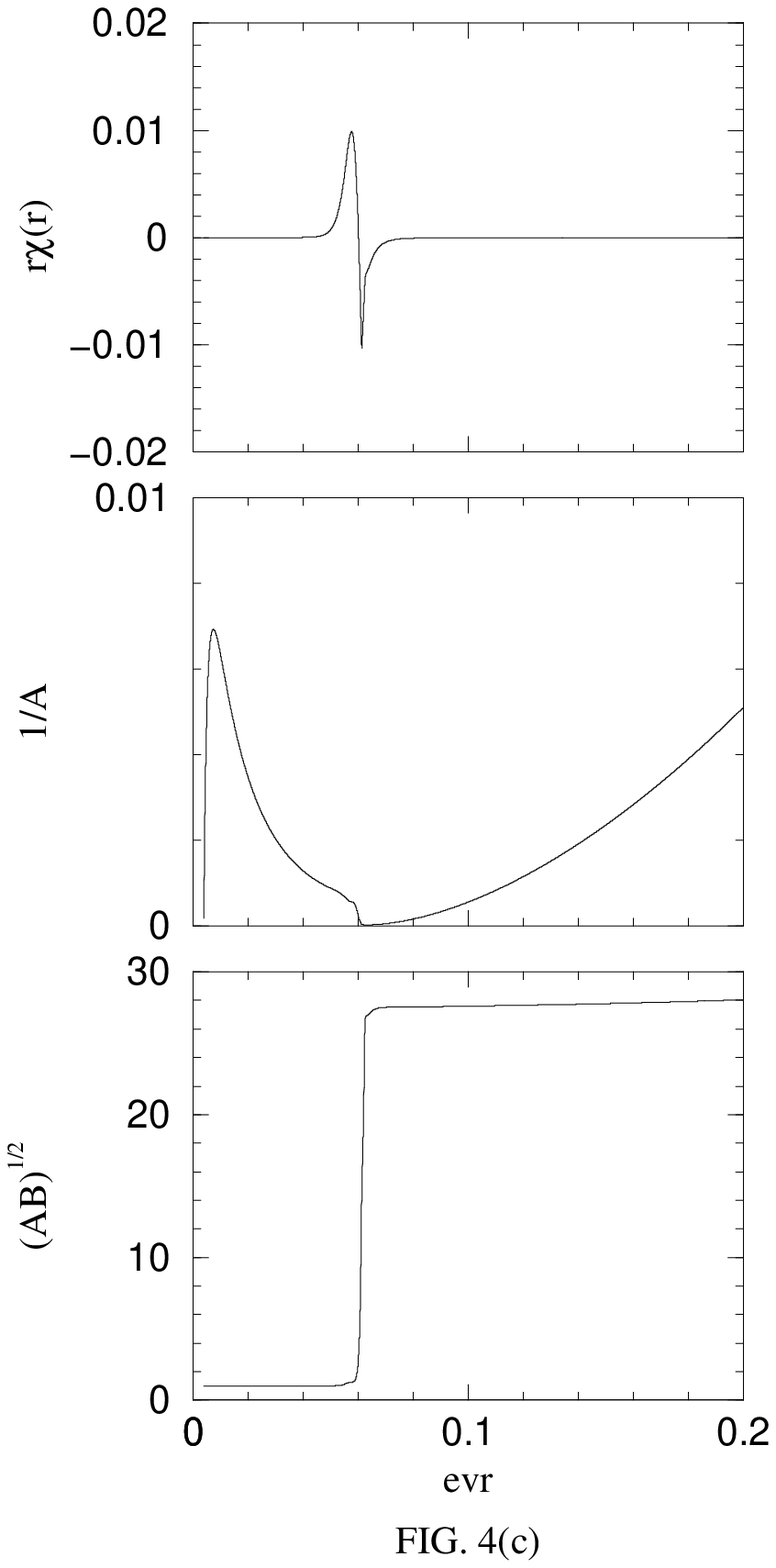
hscale=100 vscale=100 hoffset=-200 voffset=-15}{3in}{6in}\end{center}
\label{fig:t3}
\end{figure}

\setcounter{figure}{3}
\begin{figure} \begin{center}\PSbox{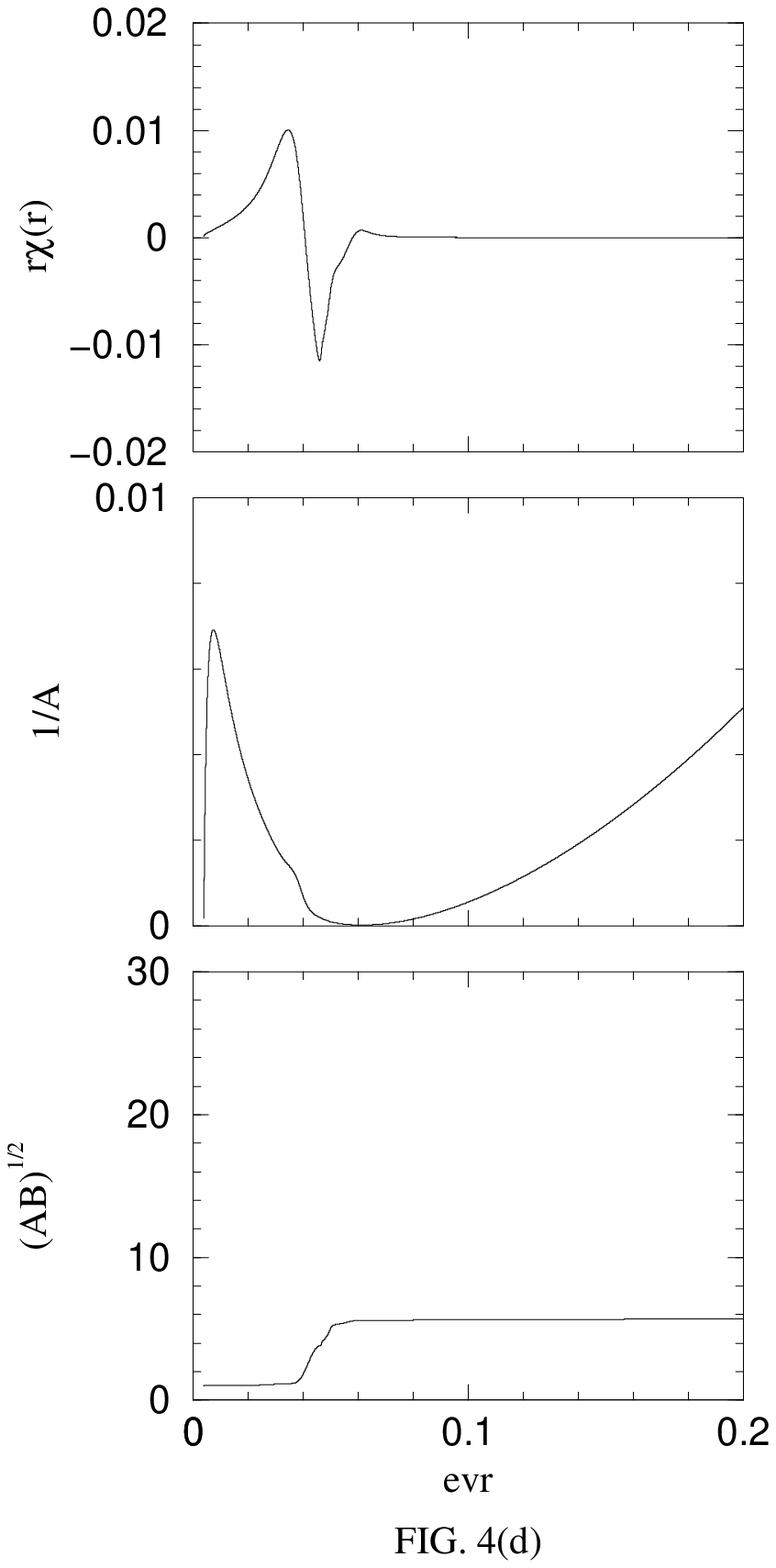
hscale=100 vscale=100 hoffset=-200 voffset=-15}{3in}{6in}\end{center}
\caption{
Evolution of $r\chi$ and the metric functions $1/A$ and
$(AB)^{1/2}$ with time.  (a) At $t = 0$, a radial Gaussian pulse in
$\chi$ is sent into a monopole with $a = 1.002$, $b = 10^6$, and a
small Schwarzschild black hole of radius $r_H = 0.004\ (ev)^{-1}$ at
its center.  The scalar field is 
coupled only to gravity and has a mass $m = 1.0\ ev$.  (b) The
configuration at $t = 56.07\ (ev)^{-1}$.  (c)
At $t = 90.72\ (ev)^{-1}$, $(1/A)_{\rm min}$ first attains the smallest
value it exhibits in this process, with $(1/A)_{\rm min} \approx
2.3 \times 10^{-5}$.  At the same time, $(AB)^{1/2}$ behind the pulse
achieves its maximum value.  (d) At $t = 113.87\ (ev)^{-1}$, the pulse
first hits the internal black hole.
}
\label{fig:t4}
\end{figure}

As required by Birkhoff's theorem, $1/A$ is is undisturbed ahead of
the pulse, but undergoes a shift as the pulse passes.  Thus, in the
region ahead of the pulse $1/A$ is the same as it was in the original
configuration, while behind the pulse it has the form corresponding to
a static monopole with a central black hole whose horizon radius
exceeds $r_H$ by an amount determined by the energy-momentum of the
infalling pulse.  These two are joined by a kink at the pulse
position.  As the pulse passes through the quasi-horizon, $1/A$
reaches its minimum.  

Similarly, the plot of $\sqrt{AB}$ appears as nearly a step function
centered at the pulse position, with the inner and outer regions
corresponding to the initial and final configurations. The jump in
$\sqrt{AB}$ at the pulse position varies with time. It reaches its
maximum when the pulse is passing through the quasi-horizon, and then
decreases.  (The resulting variation in the value of $\sqrt{AB}$ at
large $r$ may appear to violate causality, but is actually a just a
consequence of the gauge choice implicit in our choice of
coordinates.) 

As the pulse continues past the quasi-horizon into the monopole core,
the metric function $1/A$ remains static outside the pulse.  However,
$\sqrt{AB}_{\rm outside}$ decreases from its maximum value as the
pulse continues inward.  Eventually, the pulse bounces off the central
black hole (with a small amount of its energy being absorbed) and the
process reverses itself.  The pulse passes by the quasi-horizon and
retreats to infinity.  As it does so, one sees the metric variables
restore themselves to almost the original values they had before the
insertion of the pulse.

As the initial pulse amplitude is increased towards its threshold
value, the minimum value achieved by $1/A$ approaches zero, while the
maximum value of $\sqrt{AB}$ appears to grow without bound.  The time
$t_{\rm qh}$ at which the pulse passes through the quasi-horizon does
not vary appreciably.\footnote{This would not be the case if we had
fixed the normalization of $AB$ at spatial infinity; $t_{\rm qh}$
would then diverge as the threshold amplitude was approached.}

A numerical simulation will not, of course, be able to produce a
precisely extremal horizon.  In our simulations, we have been able to
adjust the pulse amplitude to make the minimum $1/A$ be as small as
$2.3 \times 10^{-5}$.  In analyzing the behavior of the solutions as
the pulse amplitude is varied, we see no indication of any singularity
as the threshold is approached.  We therefore expect that a pulse
precisely at threshold would produce a nonsingular extremal horizon.
In this case of critical collapse, the subsequent evolution of the
system would be quite similar to the subcritical case, with the pulse
continuing inward, bouncing off the central black hole, and then
retreating outward.  However, in this case the spacetime into which it
moves is causally distinct from the one where the pulse had
originated; i.e., it is a new sector of the Penrose diagram.

As a final comment, in all this analysis, dropping in pressureless
dust should give analogous results.  One can invoke Birkhoff's law
so that the metric behind the (radially thick) dust shell must be
represented by a static metric.  But one should expect the same sorts
of naked singularity behavior since this results from the interaction
of metric variables.

\section{Concluding Remarks}

In this paper we have used near-critical self-gravitating monopoles as
tools for studying the transition from a nonsingular spacetime to one
with a horizon.  By analyzing the properties of trajectories that pass
through the quasi-horizon and then emerge again, we have seen that the
observations made by an external observer vary continuously and show no
evidence of discontinuity when the critical limit is reached.  This
analysis also shows how the many causally distinct regions of the
extremal black hole spacetime naturally emerge from the simple Penrose
diagram of the subcritical monopole.   

A somewhat unexpected result from this analysis is that for the
Coulomb-type solutions the proper time required to traverse the
interior region vanishes in the critical limit.  This is closely
associated with the fact that near-critical Coulomb solutions display
naked black hole behavior; these are the first examples of
configurations without horizons that do so.  However, the absence of
this behavior in core-type solutions shows that this is not a
universal property of near-critical solutions.

Our analysis also sheds light on some aspects of black holes
themselves.  We have seen that the region bounded by the quasi-horizon
becomes effectively inaccessible to outside observers when the
solution is sufficiently close to criticality.  The interior degrees
of freedom thus become unmeasurable.  Tracing over them then leads a
naturally defined entropy that can be attributed to this
configuration.  An order of magnitude estimate of this entropy agrees
with the Hawking-Bekenstein formula for the entropy of a black hole;
it seems plausible that in the critical limit the agreement would be
precise.  These nonsingular near-critical solutions thus provide a
concrete and unambiguous framework for implementing the old idea that
black hole entropy might be understood in terms of the degrees of
freedom hidden behind the horizon.

Finally, we have argued that an extremal black hole can be produced by
allowing additional matter to fall into a near-critical monopole.  We
have illustrated this by numerical simulations.  Starting with an
initially nonsingular monopole, this leads to a zero-temperature black
hole where there had previously been no horizon at all.
Alternatively, one can start with a small black hole at the center of
the monopole.  In this latter case, a configuration with a nonzero
Hawking temperature evolves into one with $T=0$.  The existence of
these possibilities gives additional clues for, and constraints on, a
more precise formulation of the third law of black hole thermodynamics.

\acknowledgments

We wish to thank Gary Horowitz and Krishna Rajagopal for helpful
conversations.  This work was supported in part by the U.S. Department
of Energy.

\end{document}